\documentclass[a4paper,11pt]{article}
\pdfoutput=1

\usepackage{graphicx,dsfont,subfig}
\usepackage{times}
\usepackage{multirow}
\usepackage{authblk}
\usepackage{courier}
\usepackage{fullpage}
\usepackage{url}
\usepackage{verbatim} 
\usepackage{datetime}
\usepackage{mathtools}
\usepackage{xcolor}
\usepackage{bm}
\usepackage{booktabs}
\usepackage[utf8]{inputenc}
\usepackage{Definitions}
\usepackage{paralist, tabularx}

\newcommand*{\defeq}{\coloneqq}

\newcommand*{\Tall}{{T}}
\newcommand*{\Aall}{\mathcal{H}^c (T)}
\newcommand*{\Uall}{\mathcal{U}}
\newcommand*{\Qall}{\mathcal{Q}}

\newcommand*{\allTags}{\mathcal{A}}

\newcommand{\Qut}[2]{\mathcal{H}^l_{#1} (#2)}
\newcommand{\Qt}[1]{\mathcal{H}^l (#1)}

\newcommand{\A}[1]{\mathcal{H}^c (#1)}

\newcommand{\tags}[1]{\mathcal{A}_{#1}}
\newcommand{\score}[1]{s_{#1}}
\newcommand{\eug}[2]{e_{#1#2}}
\newcommand{\euallg}[1]{\bm{e}_{#1}}

\newcommand{\logNormal}[2]{\text{ln}\,\mathcal{N} (#1, #2)}

\newcommand{\paramBias}[2]{\alpha_{#1#2}}
\newcommand{\paramTrend}[2]{\mu_{#1#2}}
\newcommand{\paramKwd}[2]{k_{#1#2}}
\newcommand{\paramAllTagsBias}[1]{\bm{\alpha_{#1}}}
\newcommand{\paramAllTagsTrend}[1]{\bm{\mu_{#1}}}
\newcommand{\paramAllTagsKwd}[1]{\bm{k_{#1}}}

\newcommand{\answerTagWeights}[1]{\bm{w}_{#1}}

\newcommand*{\learningEvent}{l}
\newcommand*{\answeringEvent}{c}
\newcommand*{\paramAllBias}{{\bm{\mathcal{\alpha}}}}
\newcommand*{\paramAllTrend}{\bm{{\mathcal{\mu}}}}
\newcommand*{\paramAllKwd}{\bm{k}}

\newcommand{\xhdr}[1]{\vspace{1.5mm}\noindent{{\bf #1.}}}

\newcommand{\explain}[2]{\underset{\mathclap{\overset{\uparrow}{#2}}}{#1}}
\newcommand{\explainup}[2]{\overset{\mathclap{\underset{\downarrow}{#2}}}{#1}}

\begin{document}

\setlength{\abovedisplayskip}{2pt}
\setlength{\belowdisplayskip}{2pt}

\title{Uncovering the Dynamics of Crowdlearning \\ and the Value of Knowledge}




\author{Utkarsh Upadhyay}
\author{Isabel Valera}
\author{Manuel Gomez-Rodriguez}
\affil{Max Plank Institute for Software Systems,\\\{utkarshu, ivalera, manuelgr\}@mpi-sws.org}

\date{}

\clubpenalty=10000
\widowpenalty = 10000

\begin{small}
\maketitle
\end{small}

\begin{abstract}
Learning from the crowd has become increasingly popular in the Web and social media.
%
There is a wide variety of \emph{crowdlearning} sites in which, on the one hand, users learn
from the knowledge that other users contribute to the site,
and, on the other hand, knowledge is reviewed and curated by the same users
using assessment measures such as upvotes or likes.
%
%
%
%
%
In this paper, we present a probabilistic modeling framework of crowdlearning, which
uncovers the evolution of a user'{}s expertise over time by leveraging other users'{}
assessments of her contributions.
The model allows for both off-site and on-site learning and captures forgetting of
knowledge.
We then develop a scalable estimation method to fit the model parameters from
millions of recorded learning and contributing events.
We show the effectiveness of our model by tracing activity of $\sim$$25$ thousand users in Stack Overflow over a 4.5 year period.
We find that
%
answers with high knowledge value are rare.
Newbies and experts tend to acquire less knowledge than users in the middle range.
Prolific learners tend to be also proficient contributors that post answers with high 
knowledge value.

\end{abstract}

\noindent {\bf Keywords:} User modelling; Education; Markets and crowds; Social and information networks.

\section{Introduction}
\begin{quote}
\emph{``By learning you will teach; by teaching you will learn.''} \\
\- \hspace{45mm} ---Latin proverb
\end{quote}

Question answering (Q\&A) sites, online communities, wikis and microblogs offer unprecedented opportunities for people to learn about a wide variety of topics, acquire specialized knowledge or be up to
date with latest breaking news.
%
%
%
%
A common feature shared by most of these platforms is that knowledge is contributed by the crowd -- it is \emph{crowdsourced} -- and it is also the crowd who reviews and curates
the contributed knowledge.
For example,
in Q\&A sites, users can learn by reading answers others post to their own or similar questions;
in wikis, a set of editors write and review the content of pages in a collaborative fashion, and this content is then publicly accessible to others;
in microblogs, users post small pieces of information, which can then be assessed by other users by means of likes, shares or replies.
%
%
All of the above are examples of \emph{crowdlearning}, in which users can play the role of a \emph{learner}, a \emph{contributor}, or switch between both.
%
%
There have been many recent works on identifying experts (or potential experts) in Q\&A websites~\cite{Hanrahan:expertiseMetrics,Zhang:zscore} and microblogs~\cite{ghosh2012cognos,Pal:2011}, {as well as modeling learning in controlled settings~\cite{beck2008should,feng2005looking}}.
However, general models of crowdlearning are largely inexistent to date.
Such models are of outstanding interest since they would allow us:
\vspace{1mm}
\begin{compactitem}
\item[(i)] to better understand how people learn over time and become experts;
\item[(ii)] to identify questions with high knowledge value, which systematically help users increase their expertise;
\item[(iii)] to investigate the interplay between learners and contributors.
\end{compactitem}
\vspace{1mm}
%
%
In this paper, we propose a probabilistic generative model of crowdlearning, especially designed to fit fine-grained crowd\-learning event data~\cite{aalen2008survival}.
The key idea behind our modeling framework is simple: every time a user learns from a \emph{knowledge item} contributed by other users, she may increase her expertise and, as a consequence, her subsequent contributions
be more knowledgeable and assessed more highly by others in terms of, \eg, upvotes, likes or shares.
Thus, by jointly modeling \emph{learning events}, in which users acquire \emph{effective knowledge}, and \emph{contributing events} (in short, \emph{contributions}), in which users
contribute with their expertise to a knowledge item, our framework will reach the above mentioned goals.
In this work, we aim to measure those aspects of the learning process for which we have evidence in the observed data,
\ie, a measure of \emph{effective knowledge} that leads to measurable increase in users' \emph{effective learning}. {A general model of abstract knowledge and learning remains a challenging endeavor.}
%
%
%
%
%

In more detail, we model each user'{}s expertise as a latent stochastic process that evolves over time and think of the other users'{} assessment of her contributions as noisy samples from this stochastic process localized in time.
Moreover, this stochastic process is driven by two types of learning: off-site learning and on-site learning.
The proposed formulation also captures characteristic properties of the learning process, previously studied in the literature, such as forgetting~\cite{loftus1985evaluating} and initial expertise~\cite{Posnett12}.
We then develop an efficient parameter estimation method that finds the model parameters that maximize the likelihood of an observed set of learning and contributing events via convex optimization.
Finally, we show the effectiveness of our model by tracing learning and contributing events in data gathered from Stack Overflow over a 4.5 year period.
Our experiments reveal several interesting insights:
\vspace{1mm}
\begin{compactitem}
\item [I.] The knowledge value of items follow a log-normal distribution.

\item [II.] Users with very low or very high initial expertise, \ie, newbies and experts, tend to increase their
knowledge the least and, in contrast, users in the middle range tend to increase it the most. This suggests that the learning curve may be sigmoidal, in agreement with existing literature~\cite{Leibowitz2010}.

\item [III.] {Although there are fewer contributors than learners in absolute numbers, the distribution of knowledge in the contributions is fat tailed while the distribution of knowledge learned is heavy tailed.}

\item [IV.] Users who learn from high knowledge items are also more proficient at providing answers with high knowledge value.


\end{compactitem}
\vspace{1mm}

%
%
%

\xhdr{Related work}
Our work lies in the intersection between expert identification, learning and knowledge tracing and student modeling.

Identifying topical authorities or experts, \ie, users who provide high quality contributions, on Q\&A~\cite{Hanrahan:expertiseMetrics, Jurczyk:HITS, Pal:ExpertiseIdentification, Zhang:zscore}
and microblogging sites~\cite{ghosh2012cognos,Pal:2011}, has received a lot of attention recently.
The problem of expert finding in Q\&A sites was first studied by Zhang et~al.~\cite{Zhang:zscore}, who formulated it as a ranking problem and developed
several PageRank based methods. Shortly after, Jurczyk and Agichtein~\cite{Jurczyk:HITS} tackled the problem using link analysis techniques.
More recently, Pal and Konstan~\cite{Pal:ExpertiseIdentification} approached the problem from the perspective of supervised learning and developed Gauss\-ian
class\-ification models to distinguish between ordinary and (potential) experts users, and Hanrahan et~al.~\cite{Hanrahan:expertiseMetrics} described a method to
find experts given a specific target question.
In the context of microblogging, the problem of expert finding was first studied by Pal and Counts~\cite{Pal:2011}, who proposed a set of features for characterizing
contributors and then formulated the problem using unsupervised learn\-ing in this feature space.
Since then, Ghosh et~al.~\cite{ghosh2012cognos} mined Twitter users'{} lists to find topical authorities and Kao et al.~\cite{Kao2010} and Paulina et al.~\cite{Paulina2015} leveraged temporal
statistics on the users'{} activity to identify experts.
Finally, in the context of web search, White et al.~\cite{White2009ExpertiseWebSearch} studied how expertise influence search and Eickhoff et al.~\cite{Eickhoff2014sessionlearning} investigated
how a user can increase her expertise as she looks for \emph{procedural} and \emph{declarative} knowledge using a search engine.
%
%
{However, in contrast to our work, previous work on expert identification did not capture the evo\-lu\-tion of users'{} expertise over extended periods of time nor accounted for the knowledge value of their contributions.}

The interest in the field of modeling and measuring learn\-ing is very old and several paradigms have been developed over the last century in the experimental psychology literature~\cite{embretson2013item, means2009evaluation, norman1992psychological}. Most of the research has, however, either happened in strictly controlled environments (\ie, schools or study groups) or used centralized assessments (\eg, SAT or local testing).
%
%
The work most closely related to ours is knowledge tracing and student modeling, which has been carried on by researchers from the learning analytics, educational data mining and intelligent tutoring
communities.
In this line of work, several probabilistic models have been proposed: Bayesian knowledge tracing~\cite{corbett1994knowledge, gonzalez2013and, qiu2016modeling, yudelson2013individualized},
performance factor analysis~\cite{pavlik2009performance} and ensembles~\cite{d2011ensembling}.
However, previous work typically relies on controlled assessment and manually annotated knowledge items, even if allowing for different knowledge values per item~\cite{beck2008should}.
Only very recently, Piech et~al.~\cite{piech2015deep} solved this limitation by resorting to recurrent neural networks to model the learning of students, unfortunately, they use metrics that are
not suitable for crowdlearning.

In summary, our goal here is a general
understanding of crowdlearning dynamics, from uncovering the evolution of users'{} expertise over time and understanding the interplay between learning and contributing, to identifying
questions with a high knowledge value, which systematically help users to increase their expertise.
In contrast, previous work has focused either on identifying experts, not their expertise evolution, or modeling learning of students in controlled environment with manually annotated knowledge
items.

\section{A Crowdlearning Model}\label{sec:model}
In a crowdlearning site, users often play two different functional roles: \emph{contributors} and \emph{learners}.
In the former role, they share their knowledge on a topic (or topics) with other users within
the site; and, in the latter role, they acquire knowledge by reading what other users contributed
to the site.
Then, we can think of users'{} expertise as latent stochastic processes that evolve over time,
and think of the assessments of their contributions to the site as noisy samples from these stochastic processes localized
in time.
Here, we propose a modeling framework that uncovers the evolution of these processes by
modeling two types of learning:
\vspace{1mm}
\begin{compactitem}
\item[I.] \emph{Off-site learning}, which accounts for the knowledge that the user accumulates outside the site; and,
\item[II.] \emph{On-site learning}, which accounts for the knowledge that the user gains by reading other users'{}
contributions within the site.
\end{compactitem}
\vspace{1mm}
Next, we formulate our generative model, starting from the data it is design for.


\xhdr{Crowdlearning data} Given a crowdlearning site with a set of users $\Ucal$ and a set of learning areas (or topics) $\allTags$,
we first define a knowledge item $q$ as the smallest quantum of knowledge a user can learn from within the site.
For example, in a Q\&A site, a knowledge item corresponds to a question and its answer(s); in Twitter, it corresponds to a tweet;
and in a wiki site, it corresponds to a wiki page.
Intuitively, each knowledge item $q$ provides certain (latent) knowledge value, $\paramKwd{q}{} \in \RR^{+}$, and contains knowledge about a subset of topics
$\tags{q} \in \allTags$.
Here, we assume that knowledge is additive, \ie, $\paramKwd{q}{} =\sum_{a\in \tags{q}} \paramKwd{q}{a} = \answerTagWeights{a}^{T} \paramAllTagsKwd{q}{}$, where $\paramKwd{q}{a} \in \RR^+$ is the knowledge value contained in item $q$ about topic $a$, $\paramAllTagsKwd{q}=[\paramKwd{q}{a}]_{a \in \allTags}$, and $w_{qa} = 1$ if $a \in \tags{q}$ and $w_{qa} = 0$, otherwise. The model can be extended to non-binary weights to represent fractional presence of topics in a knowledge item~\cite{blei2003latent}.

Then, we define two types of events: \emph{learning events}, in which users acquire knowledge by reading contributions by other users, and \emph{contributing events} (or
\emph{contributions}), in which users contribute to the {crowd} by sharing their knowledge. Formally,
%
we represent each {learning event} as a triplet
  \begin{equation}
    \learningEvent~~\defeq~~(~\explain{u}{\text{\vphantom{k}user}},~~~\explain{t}{\text{time}},~~~\explainup{q}{\text{knowledge item}}~),
    \label{eqn:defQ}
  \end{equation}
which means that a user $u \in \Ucal$ \emph{learned} from knowledge item $q$ at time $t$. Here, a knowledge item $q$ may contain one or more contributions by other users. For
example, in a Q\&A site, a knowledge item corresponds to a question and its answers, typically contributed by different users. In a learning event, we do not distinguish the knowledge provided
by individual contributions, but instead, consider the knowledge of the item as a whole. Moreover, note that, if the knowledge value of an item is zero, the learning event
will not increase the expertise of the learner.
Then, we denote the history of lear\-ning events associated to user $u$ up to time $t$ by $\Qut{u}{t} = \bigcup_{i: t_i < t} \left\{ \learningEvent_{i} \mid u_i = u \right\}$, and the history of learning
events in the whole crowdlearning site up to time $t$ by $\Qt{t} = \bigcup_{i: t_i < t} \left\{ \learningEvent_{i} \right\}$.

%
Similarly, we represent each contribution as a quadruplet
  \begin{equation}
    \answeringEvent~~\defeq~~(~\explain{u}{\text{\vphantom{k}user}},~~~\explain{t}{\text{time}},~~~\explainup{q}{\text{knowledge item}},~~~\explain{\score{}}{\text{score}}~),
    \label{eqn:defA}
  \end{equation}
which means that a user $u \in \Ucal$ contributed to a knowledge item $q$ at time $t$, and other users assigned a score $\score{}$ to her contribution. For example, in a Q\&A site, this may be the number of upvotes an answer receives. We gather the history of contributions in the whole crowdlearning site up to time
$t$ by $\A{t} = \bigcup_{i:t_i < t} \left\{ \answeringEvent_i \right\}$, and the history of contributions and learning events up to time $t$ by $\Hcal(t) = \A{t} \bigcup \Qt{t}$.

\xhdr{Crowdlearning generative process} We represent each user'{}s expertise as a multidimensional (latent) stochastic process $\euallg{u}^{*}(t)$, in which the $a$-th entry, $\eug{u}{a}^{*}(t) \in \RR^+$, represents the user $u$'{}s expertise on
topic $a$ at time $t$. Here, the sign $^{*}$ means that the expertise $\eug{u}{a}^{*}(t)$ depends on her learning history $\Qut{u}{t}$. Then, every time a user $u$ contributes to a knowledge item $q$
at time $t$, we draw the contribution's score from a distribution $p(s | \tags{q}, \euallg{u}^{*}(t))$.
%
%
%
Further, we represent the times of the learning and contributing events within the site by two sets of counting processes, denoted by two vectors $\Nb^{l}(t)$ and $\Nb^{c}(t)$, in which the $u$-th entries, $N^{l}_{u}(t)$
and $N^{c}_{u}(t)$, count the number of times user $u$ learned from and contributed to the crowdlearning site up to time $t$.
Then, we can characterize these counting processes using their corresponding intensities as
\begin{equation*}
\EE[d\Nb^{l}(t)\, |\,  \Hcal(t)] = \lambdab^{l}(t) \, dt \enspace \mbox{and} \enspace \EE[d\Nb^{c}(t)\, |\,  \Hcal(t)] = \lambdab^{c}(t) \, dt
\end{equation*}
where $d\Nb^{l}(t):=[dN^{l}_{u}(t)]_{u \in \Ucal}$ and $d\Nb^{c}(t):=[dN^{c}_{u}(t)]_{u \in \Ucal}$ denote the number of learning and contributing events in the window $[t, t+dt)$ and
$\lambdab^{l}(t) := [\lambda_{u}^{l}(t)]_{u \in \Ucal}$ and $\lambdab^{c}(t) := [\lambda_{u}^{c}(t)]_{u \in \Ucal}$ denote the vector of intensities associated to all the users.
Here, there is a wide variety of intensity functions one can choose from~\cite{aalen2008survival}. However, modeling the times of learning and contributing events is not the main focus of this work -- we refer the reader
to the growing literature on social activity modeling using point processes~\cite{farajtabar2014activity,competing15icdm,zhou2013learning}.
Next, we specify the functional form of each user'{}s expertise $\euallg{u}^{*}(t)$ and the score distribution $p(s | \tags{q}, \euallg{u}^{*}(t))$.

\xhdr{Stochastic process for expertise}
The expertise $\eug{u}{a}^{*}(t)$ of a user $u$ on a topic $a$ at time $t$ takes the following form:
  \begin{align*}     
    \eug{u}{a}^{*}(t) & \defeq \hspace{2mm}
      \overbrace{\paramBias{u}{a}}^{\mathclap{\text{initial expertise}}} \hspace{2mm}+  \underbrace{\paramTrend{u}{a} \cdot t}_{\text{off-site learning}} +
        \overbrace{\sum_{i : q_i \in \Qut{u}{t} } \paramKwd{q_i}{a} \cdot \kappa_{\omega}\left( t - t_i \right)}^{\text{on-site learning}}
  \end{align*}
where the first term, $\paramBias{u}{a} \in \RR^+$, models the initial expertise of user $u$ on a topic $a$ when she joined the crowdlearning site; the second term, $ \paramTrend{u}{a} \in \RR^+$, assumes a linear trend
for the off-site learning process as a first order approximation\footnote{\scriptsize Several other shapes for the learning curve have been proposed in Heathcote, et~al.~\cite{heathcote2000power}. However, we chose the linear form for its simplicity and ease in model estimation, as suggested by Skinner~\cite{skinner2010applied}.}; and, the third term models the knowledge a user acquires by means of learning events within the crowdlearning site.
%
Here, $\kappa_{\omega}(t)$ is a nonnegative kernel function that models the rate at which users forget the knowledge they learn from knowledge items.
%
Following previous work on the psychology literature~\cite{averell2011form,loftus1985evaluating}, which argues that people forget at an exponential rate, we opt for an exponential kernel $\kappa_{\omega}(t) := \exp(-\omega t)\II(t \geq 0)$.
However, our model estimation method does not depend on this particular choice.
%
%
%
%

For compactness, we write each user'{}s expertise as a row vector of length $|\allTags|$, \ie,
\begin{align}   \label{eq:expertiseVectorForm}
  \euallg{u}^{*}(t) &= \paramAllTagsBias{u} + \paramAllTagsTrend{u} \cdot t +
  \sum_{i: q_i \in \Qut{u}{t}} \paramAllTagsKwd{q_i} \cdot \kappa_{\omega}(t - t_i)
\end{align}
where $\paramAllTagsBias{u}=[\paramBias{u}{a}]_{a \in \allTags}$, $\paramAllTagsTrend{u}=[\paramTrend{u}{a}]_{a \in \allTags}$ and
$\paramAllTagsKwd{q_i}=[\paramKwd{q_i}{a}]_{a \in \allTags}$. Here, by definition, $\paramKwd{q_i}{a}=0$ if $a \notin \tags{q_i}$.
Then, we can gather the model parameters for all users in three matrices $\paramAllBias$, $\paramAllTrend$ and $\paramAllKwd$ with sizes  $|\Uall| \times |\allTags|$, $|\Uall| \times |\allTags|$ and
$|\Qall| \times |\allTags|$.
%

\xhdr{Score distribution}
Given a contribution $\answeringEvent = \left( u, t, q, \score{} \right)$, the particular choice of score distribution $p(s | \tags{q}, \euallg{u}^{*}(t))$ depends on the observed data.
In this work, we consider discrete non-negative scores, $s \in \NN$, which fit well several scenarios of interest. For example, in Stack Overflow, scores may correspond to the number of upvotes that answers receive; in
Twitter, to the number of likes or retweets that tweets receive; and, in Pinterest, to the repins that a pin receives.
A natural choice in such cases is the Poisson distribution:
\begin{equation} \label{eq:scoreDistribution}
p(s | \tags{q}, \euallg{u}^{*}(t)) \sim
\text{Poisson} \left( \frac{\answerTagWeights{q}^{T} \euallg{u}^{*}(t)}{\answerTagWeights{q}^{T}{\mathbf{1}}} \right),
\end{equation}
Here, $\mathbf{1}$ is a column vector of ones with length $|\allTags|$.
With this choice, the average of the score distribution is simply the average expertise of user $u$ at time $t$ across the topics $\tags{q}$ the knowledge
item $q$ is about.
Moreover, the greater the expertise of a user, the greater the scores given by other users to her contributions, as one may expect in real-world data.

Note that depending on the recorded data, we could choose a different score distribution, \eg, for continuous assessments like time elapsed between the question and the answer, one may choose a continuous distribution.
Our model estimation method in Section~\ref{sec:estimation} can be easily adapted to any distribution that is jointly log-concave with respect to the model parameters $\paramAllBias$,
$\paramAllTrend$ and $\paramAllKwd$.

\xhdr{Efficient Parameter Estimation}\label{sec:estimation}
Given a collection of learning and contributing events, $\Qt{T}$ and $ \A{T}$, recorded during a time period $[0, T)$, we find the optimal model parameters $\paramAllBias$, $\paramAllTrend$ and
$\paramAllKwd$ by solving the following maximum likelihood estimation problem:
\begin{equation} \label{eqn:optimizationProb}
        \underset{\paramAllBias \geq 0, \paramAllTrend \geq 0, \paramAllKwd \geq 0}{\text{maximize}} \Lcal(\paramAllBias, \paramAllTrend, \paramAllKwd),
\end{equation}
where we compute the log-likelihood $\Lcal(\paramAllBias, \paramAllTrend, \paramAllKwd)$ using Eq.~\ref{eq:expertiseVectorForm} and Eq.~\ref{eq:scoreDistribution}, \ie,
\begin{equation}
  \Lcal(\paramAllBias, \paramAllTrend, \paramAllKwd) = \sum_{\substack{(u, t, q, s) \\ \in \Aall}}
        s \cdot \log{\left( \frac{\answerTagWeights{q}^{T} \euallg{u}^{*} (t)}{\answerTagWeights{q}^{T}{\mathbf{1}}} \right)} -
         \frac{\answerTagWeights{q}^{T} \euallg{u}^{*} (t)}{\answerTagWeights{q}^{T}{\mathbf{1}}}.
\label{eqn:LL}
\end{equation}

\begin{figure}[t]
\centering
\subfloat[LE per question]{\includegraphics[width=0.25\textwidth]{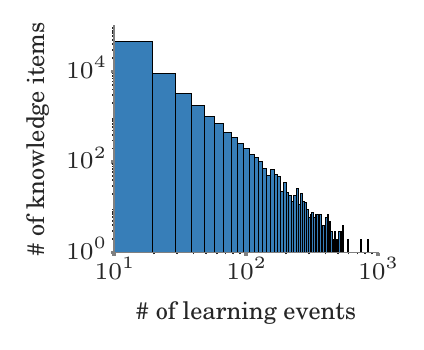}\label{fig:learning-events-per-question}}
\subfloat[LE per user]{\includegraphics[width=0.25\textwidth]{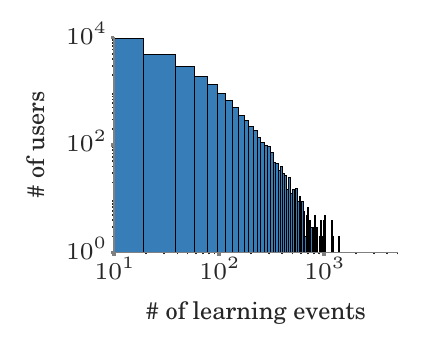}\label{fig:learning-events-per-user}}
\subfloat[LET per user]{\includegraphics[width=0.25\textwidth]{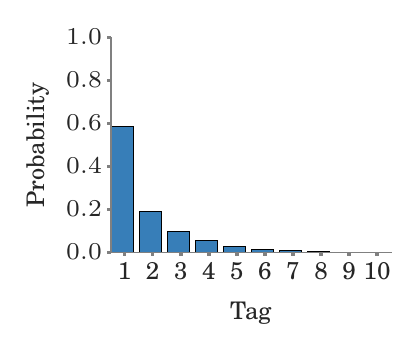}\label{fig:questions-tags-per-user}}
\subfloat[CET per user]{\includegraphics[width=0.25\textwidth]{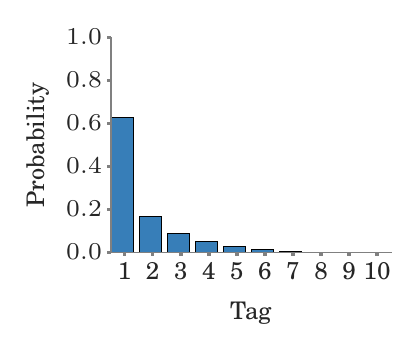}\label{fig:answers-tags-per-user}}
\caption{Statistics of learning events (LE), tags of learning events (LET) and tags of contributing events (CET) in the Stack Overflow dataset. In Panels (c) and (d), the $x$-axis denotes the tag index in order of popularity for each user.}\label{fig:stats}
\end{figure}

Since $\euallg{u}^{*}(t)$ is linear in the model parameters $\paramAllBias, \paramAllTrend$ and $\paramAllKwd$,
the function $\log x - x$ is concave, and a composition of a linear function with a linear combination
of concave functions is concave,
the optimization problem above is jointly convex in $\paramAllBias$, $\paramAllTrend$, and $\paramAllKwd$.
As a consequence, the global optimum can be efficiently found by many algorithms.
In practice, the limited memory BFGS with bounded variables (L-BFGS-B) algorithm~\cite{Zhu:1997} worked best for our problem.

\xhdr{Remarks} In this work, we are measuring \emph{effective learning},
which accounts for the ability of a user to get better assessment of her
posts, and \emph{effective knowledge}, which accounts for the gain in this
ability that learning from a knowledge item causes. Making these quantities
correspond to real-life expertise and knowledge value on a crowdlearning
website requires careful mapping from the features on that website to learning
events and scores.

Moreover, using our model, one can only measure learning and knowledge if there is overlap between the topics of a user's learning and contributing events. Therefore, there is a trade-off between the granularity of the topics chosen and the amount of data available for inference: increasing the granularity ensures accurate mappings between learning and contributing events, but reduces the amount of data available to learn the model parameters. We discuss this further in Section~\ref{sec:discussion}.


\begin{figure}[t]
\centering
\subfloat[Expertise evolution]{\includegraphics[width=.40\textwidth]{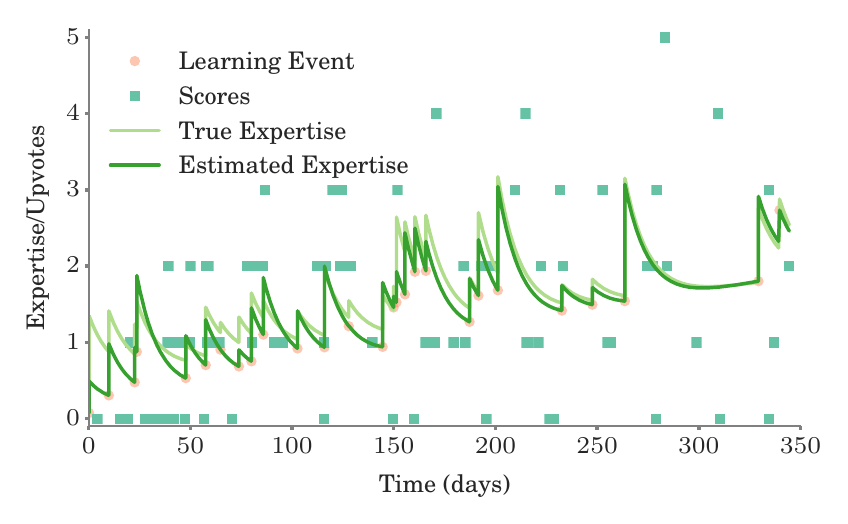}\label{fig:learning-trajectory-synthetic}}%
\subfloat[$1$ tag]{\makebox[0.30\textwidth][c]{\includegraphics[width=0.3\textwidth]{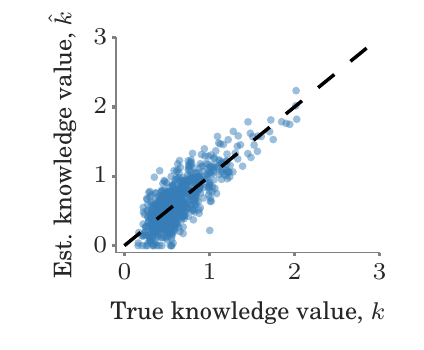}}}\label{fig:knowledge-item-1-tag-synthetic}%
\subfloat[$10$ tags]{\makebox[0.30\textwidth][c]{\includegraphics[width=0.3\textwidth]{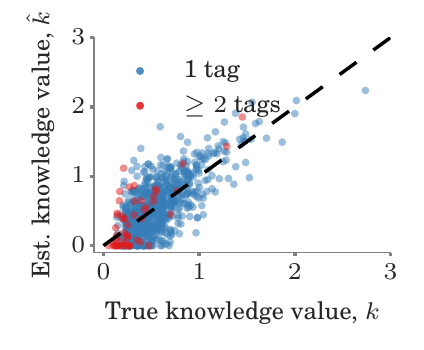}}}\label{fig:knowledge-item-10-tag-synthetic}

\caption{In (a), estimated and true expertise evolution for a user, picked at random, in the $1$-tag synthetic dataset.
In (b) \& (c), estimated (y-axis) against true (x-axis) knowledge item values. Each point corresponds to a knowledge item variable, and the line defined by $x = y$ corresponds to zero estimation error. Our estimation
method achieves Spearman'{}s correlations $\rho_{\mbox{\scriptsize 1-tag}} = 0.74$ and $\rho_{\mbox{\scriptsize 10-tag}} = 0.64$.}\label{fig:scatter-synthetic-items}

\end{figure}

\begin{figure}[t]
\centering
\subfloat[$\paramAllTrend$ ($1$-tag)]{\makebox[0.3\textwidth][c]{\includegraphics[width=.3\textwidth]{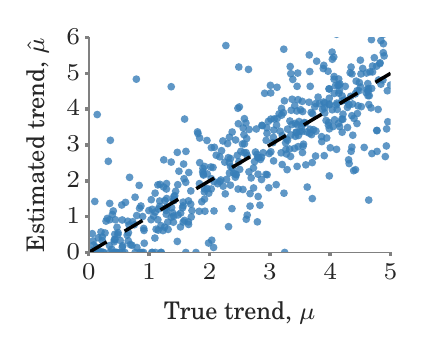}}\label{fig:trend-1-tag-synthetic}} \hspace{0.1mm}
\subfloat[$\paramAllBias$ ($1$-tag)]{\makebox[0.3\textwidth][c]{\includegraphics[width=.3\textwidth]{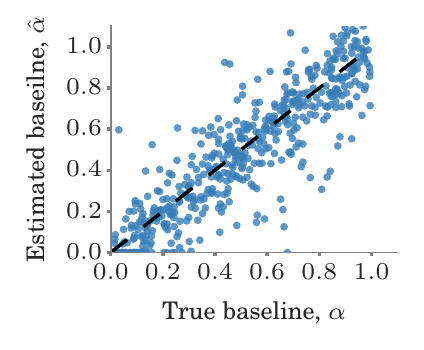}}\label{fig:baseline-1-tag-synthetic}}
\caption{Estimated (y-axis) against true (x-axis) model parameters for the $1$-tag synthetic datasets. Each point corresponds to a user'{}s (a) trend $\mu_u$ or (b) baseline $\alpha_u$
variable, and the line defined by $x = y$ corresponds to zero estimation error. Our estimation method achieves a Spearman'{}s correlation $\rho_{\mu} = 0.82$
and $\rho_{\alpha} = 0.89$. The results for $10$-tag synthetic datasets are qualitatively similar.
}
\label{fig:scatter-synthetic-trend-bias}
\end{figure}

%
%
\section{Experiments on synthetic data} \label{sec:synthetic-experiments}
In this section, we first show that our model estimation method can accurately recover the true model parameters from learning and contributing events
generated under \emph{realistic} conditions.
We then show that, as long as
there are a sufficient number of contributions \emph{per} learning event,  the estimation becomes more accurate as we feed more events into the estimation procedure.
Finally, we show that our estimation method can easily scale up to millions of users, knowledge items, and learning and contributing events.

\begin{figure*}[t]
\centering
\subfloat[CR vs.\ learning events]{\includegraphics[width=0.3\textwidth]{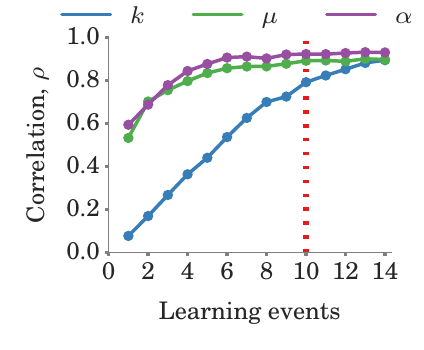}\label{fig:correlation-learning-events-synthetic}}
\subfloat[CR vs.\ contributions]{\includegraphics[width=.3\textwidth]{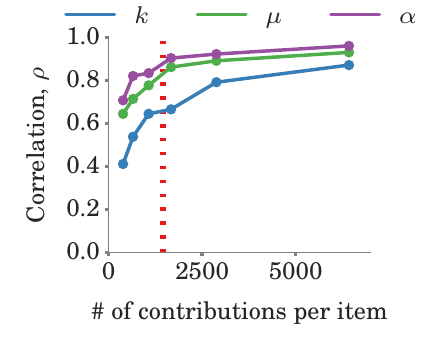}\label{fig:correlation-contributing-events-synthetic}}
\subfloat[RMSE ($\mu$) vs.\ contributions]{\includegraphics[width=.3\textwidth]{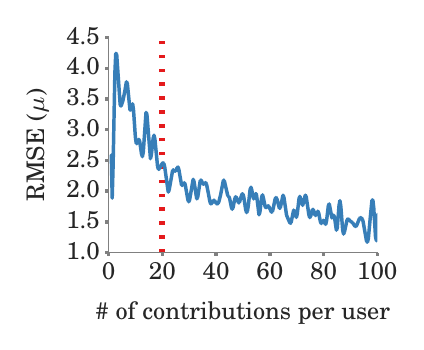}\label{fig:trend-rmse-with-contributions}}

\caption{Estimated against true model parameters for the $1$-tag synthetic dataset. Panels (a) and (b) show the correlation (CR) between the estimated and true model parameters against number of learning events and
median number of contributions per knowledge item, respectively. Panel (c) shows the RMSE for the estimated trends, $\paramAllTrend$, against number of contributed events per user.
{In Panel (a), the number of contributing events is $255{,}000$ and the red dotted line shows the threshold (10) we chose for the learning events per knowledge item in our dataset (see Section~\ref{sec:real-experiments}).
In Panels (b) and (c), the number of learning events is always $13{,}000$ and the red dotted lines show the median number of contributions per knowledge item and the minimum number of contributions per user
in the experiments on our dataset, respectively (see Section~\ref{sec:real-experiments}).}}
\label{fig:correlation-learning-contributing-events-synthetic}
\end{figure*}

\xhdr{Experimental setup}
We carefully craft an experimental setup to closely mimic some of the empirical patterns observed in real crowdlearning data, as given by
Figure~\ref{fig:stats}. Here, for simplicity, we assume the topics associated to each knowledge item are specified by means of tags.

Given a set of users and knowledge items, we draw the users'{} offsite learning rates $\{ \paramTrend{u}{a} \}$ and initial expertise $\{ \paramBias{u}{a} \}$ from $U(0, 5)$ and $U(0, 1)$, and the knowledge value of the items from the rescaled log-normal distribution $0.05 \times \logNormal{0}{1}$. These choices ensure that the distribution of scores which users receive resembles the distribution in real data. We set the users'{} forgetting decay rate to $\omega = (11.6\text{ days})^{-1}$, such that 50\% of the knowledge is forgotten roughly after the first week, and
assume that the intensities of both users'{} learning and contributing events are (homogeneous) Poisson processes.
We denote the total simulation time by $T$.
We set each user'{}s learning event rate to $T/n$, where $n$ is drawn from a log-normal distribution, so that the number of events
per user fits well the empirical distribution (see Figure~\ref{fig:learning-events-per-user}), and each user'{}s
contributing rate to $T/m$, where $m$ is drawn from an uniform distribution for easy control.
Finally, for each user, we shuffle the tag labels and set her tag learning propensity, defined as the probability that she up-votes a knowledge
item with a given tag, and her tag contributing propensity, defined as the probability that she contributes to a knowledge item with a given tag, using
the empirical distributions (see Figures~\ref{fig:questions-tags-per-user} and~\ref{fig:answers-tags-per-user}).

Then, we generate learning and contributing events as follows\-.
First, we generate the timings of each user'{}s lear\-ning events by drawing samples from the corresponding Poisson\- process, and assign
each learning event to a knowledge item such that the user'{}s tag learning propensity is satisfied.
%
%
Then, we generate the timings of each user'{}s  contributions by drawing samples from the corresponding Poisson process,
assign each contributing event to a knowledge item such that the user'{}s tag contributing propensity is satisfied,
and draw the quality score from a Poisson distribution that depends on the user'{}s expertise on the item tags at the time of the event, as
given by Eq.~\eqref{eq:scoreDistribution}.
%
%
Unless explicitly stated, we only consider knowledge items with at least $10$ associated learning events.
Given this data, our goal is to find the knowledge value of the items users learned from, as well as
the users'{} offsite learning rates and initial expertise by solving the maximum likelihood estimation problem defined in Eq.~\eqref{eqn:optimizationProb}.

\xhdr{Parameter estimation accuracy}
%
We evaluate the accuracy of our model estimation procedure across all users and knowledge items for a $1$- and $10$-tag dataset with $\sim$$800$ knowledge items. Figure~\ref{fig:scatter-synthetic-items} summarizes the
results for the estimation of the knowledge item values by means of two scatter plots.
In all cases, we find that points lie close to the line $x = y$, \ie, their estimation error is close to zero. We also observe that the estimation of knowledge items in the $10$-tag dataset is more challenging than in the
$1$-tag dataset.
Additionally, Figure~\ref{fig:scatter-synthetic-trend-bias} summarizes the results for the estimation of the user'{}s expertise baseline and trend variable for the $1$-tag dataset using scatter plots. Results for the $10$-tag dataset are qualitatively similar although the estimation is more challenging.
In particular, if we look at the estimation of the knowledge values, trends and baseline for the $1$-tag dataset, our estimation method achieves a Spearman'{}s correlations $\rho_{k} = 0.74$, $\rho_{\mu} = 0.82$ and $\rho_{\alpha} = 0.89$
while, for the $10$-tag dataset, it achieves  $\rho_{k} = 0.64$, $\rho_{\mu} = 0.76$ and $\rho_{\alpha} = 0.81$.
This is most likely due to the mix\-ing of tag knowledge variables within the same knowledge item, \ie, the linear combination of knowledge variables in Eq.~\eqref{eqn:LL}.
\begin{figure}[t]
\centering
\subfloat[RT vs.\ learning events]{\makebox[0.40\textwidth][c]{\includegraphics[width=.33\textwidth]{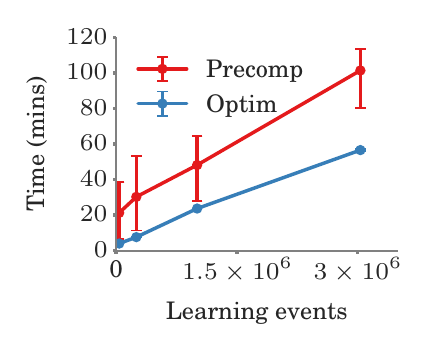}}\label{fig:run-time-learning-events-synthetic}} 
\subfloat[RT vs.\ contributions]{\makebox[0.40\textwidth][c]{\includegraphics[width=.33\textwidth]{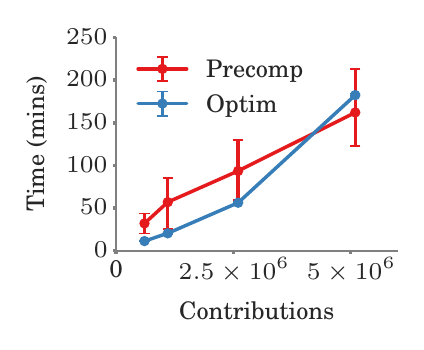}}\label{fig:run-time-contributing-events-synthetic}} \\
\caption{Running time (RT) of our model estimation method. In Panel (a), we consider $\sim$$2$ million contributing events while varying number of learning events (and knowledge items); in Panel (b), we consider $\sim$$1.8$ million learning events while varying the number of contributions (per learner). For pre-processing, we
used ten machines with $48$ cores and, for the optimization itself, we used a single machine with $48$ cores.
The memory requirements were below $16$ GB at all points of the pre-processing and optimization.}
\label{fig:run-time-learning-contributing-events-synthetic}
\end{figure}

\xhdr{Parameter estimation accuracy vs.\ number of learn\-ing events}
In our model, we can think of learning events as measurements of the amount of knowledge in a knowledge item, which are accumulated over time in the users'{} expertise, and of contributing events as noisy
samples of the users'{} expertise at particular points in time.
Therefore, intuitively, the more users learn from a knowledge item the easier it should become to accurately estimate their associated knowledge value, as long as these users also contribute to other knowledge items with overlapping topics.
Figure~\ref{fig:correlation-learning-events-synthetic} confirms this intuition by showing the Spearman'{}s correlation against minimum number of learning events per knowledge item in a $1$-tag dataset with
$255{,}000$ contributions.

\xhdr{Parameter estimation accuracy vs.\ number of contributing events}
As pointed out above, we can think of the score of contributing events as noisy samples of users'{} expertise at particular points in time. Therefore, one may expect the accuracy of our model parameter
estimation to improve as the number of contributions increases, due to a more fine-grained sampling of each user'{}s expertise.
Fig\-ure~\ref{fig:correlation-contributing-events-synthetic} gives empirical evidence that this indeed happens, by showing the Spearman'{}s correlation against average number of answers \emph{per}
learning event in a $1$-tag dataset with $13{,}000$ learning events. Figure~\ref{fig:trend-rmse-with-contributions} shows how the RMSE of the estimation of $\paramAllTrend$ decreases as the number of contributions made by the user increases.

\xhdr{Scalability of parameter estimation}
Crowd-learning sites such as Stack Overflow or AskReddit are rapidly increasing their number of active users, questions and answers. For example, Stack Overflow recently crossed the $\sim$10 million questions
mark\footnote{\scriptsize \url{http://meta.stackoverflow.com/questions/303045/10-million-questions}}.
The pre-computation of all coefficients in Eq.~\eqref{eqn:LL}, which is the running time bottleneck, can be readily parallelized.
Figure~\ref{fig:run-time-learning-contributing-events-synthetic} shows that our model estimation method easily handles up to millions of learning events and contributing events, and scales almost linearly with
the number of learning events and contributions.

Thus, it should be possible to scale up our estimation method even further.

\section{Experiments on Real data}\label{sec:real-experiments}
In this section, we apply our model estimation method to a large-scale crowdlearning dataset from Stack Overflow.
%
First, we evaluate our model quantitatively by means of a prediction task: given two different answers to a question, predict which one will receive a higher score.
Then, we discuss the distribution of the knowledge values and the effect of the kernel parameter on the estimation, identify different types of learners and derive
insights into their main characteristics.
Finally, we study the interplay between learners and contributors in crowdlearning sites and investigate to which extent users switch between
lear\-ning and contributing over time.

\xhdr{Data description}
Our Stack Overflow dataset comprises $\sim$$8$ million questions, $\sim$$13.7$ million answers, and $\sim$$47.2$ million upvotes. These questions and answers were posted by $\sim$$1.9$ million
users during a six year period, from the site'{}s inception on July 31, 2008 to September 14, 2014.
Importantly, for each upvote, our dataset contains its associated user identity, question or answer identity and timestamp\footnote{\scriptsize Stack Overflow generously gave us access to these additional metadata, which allows us to readily fit our model.}.
We discard the events which happened before 2010-01-01 (before the site had fully matured) and after 2014-06-01 (the extent of the data-dump {we had access to}).
Whenever in the data a user \emph{upvotes} (\emph{writes}) an answer, we record it as a learning (contributing) event involving the user and the knowledge item
containing the answer.
Moreover, we select the number of upvotes a user's answer received in the first week after posting it as the score of the contribution; downvotes were discarded because
they constitute less than 3\% of total votes cast. Here, we consider only the first week of voting to prevent old contributions from gaining an unfair advantage as they
have more time to accumulate upvotes.
%
%
Figure~\ref{fig:stats} provides general statistics on learning and contributing events and tags usage. We find that the learning events per user (per question) follow a log-normal (power-law) distribution. As shown, the tag usage is highly skewed
towards few tags; most users contribute and learn only from their favorite tags.
%
%
%
%
%

%
%
%
\xhdr{Data preprocessing}
%
%
%
In Section~\ref{sec:synthetic-experiments}, we have shown that the accuracy of our estimation method depends dramatically on the number of learning and contributing events per question and user (refer to
Figure~\ref{fig:correlation-learning-contributing-events-synthetic}).
As a consequence, we can only expect our model estimation method to provide reliable and accurate results in real data if the data we start with contains enough learning and contributing events per question and user.
To this aim, we carefully pre-process our large-scale dataset of learning and contributing events. We only consider:
\vspace{1mm}
\begin{compactitem}
\item[(i)] Knowledge items with more than $10$ associated learning events, which corresponds to a correlation value $\ge 0.8$ between true and estimated knowledge parameters in synthetic data, as shown in
Figure~\ref{fig:correlation-learning-events-synthetic}.
%
\item[(ii)] Users that contribute (answer) more than $20$ times in at least 10 unique months, which corresponds to a RMSE value $\le 2$ for the estimated users'{} baseline and trend parameters in
synthetic data, as shown in Figure~\ref{fig:trend-rmse-with-contributions}; and,
\item[(iii)] Top 10 tags in terms of number of learning events in the recorded data (\ie, \texttt{java}, \texttt{c\#}, \texttt{javascript}, \texttt{php}, \texttt{android}, \texttt{jquery}, \texttt{python}, \texttt{html}, \texttt{c++}, and \texttt{mysql}).
\end{compactitem}
\vspace{1mm}
{After these preprocessing steps, our dataset consists of $\sim$25 thousand users who learn from $\sim$66 thousand knowledge items by means of $\sim$$1.4$ million learning events, and contribute to {$\sim$2.5 million} knowledge items, by means of $\sim$$3.8$ million contributing events.}
Then, we correct for the overall decreasing trend on number of upvotes per answer over time\footnote{\scriptsize The number of upvotes per answer decreases over time because the number of answers grows
at a faster rate than the number of learners.} and since, for each knowledge item, most learning events occur\- after all the contributions (answers) to the knowledge item took place, we assume its knowledge value to be constant. {We use the first event of each user in our dataset as as an estimate of her joining time.}

Finally, we would like to highlight that the preprocessing steps above do not aim to reduce the size of the original dataset but to increase the accuracy of our estimated model parameters and the reliability of our
derived qualitative insights --- our model estimation method does easily scale to millions of learning and contributing events. 
In this case, the pre-processing of the raw data using five machines with $48$ cores each took $\sim$$30$ minutes and our estimation method, implemented using the Intel MKL libraries, took $\sim$$11.5$ hours on
a single machine with $48$ cores.
The memory requirements were below $16$ GB at all points of the pre-processing and optimization.
\begin{table}
\centering
\begin{tabular}{crrr}
\toprule
Score difference & \# of pairs & Off-site only & Our model \\
\midrule
$\ge 1.0$ & 31,639 & 52.5\% & \textbf{61.9\%}\\
$\ge 2.0$ & 19,253 & 52.9\% & \textbf{64.8\%}\\
$\ge 3.0$ & 10,804 & 53.2\% & \textbf{67.0\%}\\
$\ge 4.0$ & 5,910 & 53.7\% & \textbf{70.7\%}\\
$\ge 5.0$ & 3,250 & 55.0\% & \textbf{71.6\%}\\
$\ge 6.0$ & 1,935 & 56.0\% & \textbf{73.3\%}\\
$\ge 7.0$ & 1,159 & 56.8\% & \textbf{73.8\%}\\
\bottomrule
\end{tabular}
\caption{Performance of our model against a linear baseline model at predicting which one of two answers to a question will receive a higher score.
As the difference in score between the answers (and, hence, the users'{} expertise) increases, the competitive advantage of our model becomes more
pronounced.}
%
\label{tab:evaluation}
\end{table}

\xhdr{Quantitative evaluation}
We evaluate our model quantitatively by means of the following prediction task: given two different answers to a question, predict which one will receive a higher score, \ie, more
number of upvotes in the week after posting it.

\vspace{1mm} \emph{--- Experimental setup:}
We train our model using the first 80\% of the answers provided by each learner, as well as the learning events which occurred before them.
Then, we match pairs of answers to the same questions from the remaining 20\% and predict which one will receive a higher score. Here, we only consider questions with pairs of
answers provided by users from our dataset such that their scores differ by at least one upvote.
There are $\sim$32 thousand such pairs in our dataset.
Finally, we compare our model against a baseline linear model which only accounts for off-site learning to show the benefits of including knowledge item variables.

\vspace{1mm} \emph{--- Results:}
Table~\ref{tab:evaluation} compares the performance of our model against the baseline model as the difference in score between the answers (and, hence, the users'{}
expertise) increases.
Our model consistently outperforms the baseline for any score difference and the competitive advantage becomes more pronounced as the score difference increases,
reaching $>$$73$\% accuracy when the score difference is $\geq$$6$.



%
\begin{figure}[t]
\centering
\subfloat[Overall knowledge values]{\makebox[0.35\textwidth][c]{\includegraphics[width=0.35\textwidth]{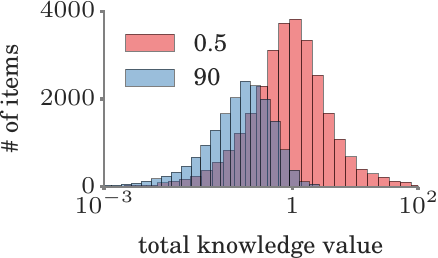}\label{fig:knowledge-item-distribution}}}
\subfloat[Fraction of upvotes leading to learning]{\makebox[0.35\textwidth][c]{\includegraphics[width=0.35\textwidth]{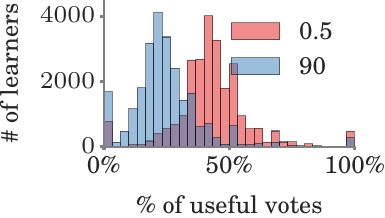}}\label{fig:useful-upvotes}}
\caption{Estimated knowledge values for knowledge items and \emph{useful upvotes} for two different kernel parameters, with \emph{half-life} 0.5 days (12 hours) and 90 days.
Panel (a) shows the distribution of knowledge value per knowledge item follows a log-normal distribution with longer half-life leading to smaller knowledge values and higher sparsity.
Panel (b) shows what fraction of upvotes were useful (\ie, led to learning) per learner: higher half-life leads to higher sparsity, which leads to fewer fraction of upvotes causing effective learning.
%
%
%
}
\end{figure}

\xhdr{Knowledge value and forgetting rate}
In this section, we leverage our model to give insights on the knowledge values across items in Stack Overflow for different forgetting rates, \ie, the kernel decay
parameter $\omega$. We express the kernel decay parameter $\omega$ in units of \emph{half-life} in days, \ie, the time to forget 50\% of the knowledge in an item.
Figure~\ref{fig:knowledge-item-distribution} shows the distribution of estimated knowledge value across knowledge items for two kernel parameters with half-life $0{.}5$
days and $90$ days. We find several interesting patterns.
Knowledge values in both settings follow a log-normal distribution, in which $\sim$$10$\% of the items account for $\sim$$75$\% of the overall knowledge.
However, while for a half-life of $0{.}5$ days, $\sim$$53$\% of the knowledge items do not contribute knowledge, this fraction increases to $\sim$$70$\% for $90$
days.
A potential explanation for this difference is that, by increasing the half-life, a knowledge item must show evidence of \emph{effective learning} over longer stretches
of time to contribute knowledge and this happens more rarely.
As a consequence, a smaller fraction of upvotes lead to effective learning (\ie, being \emph{useful}) when the half-life is high, as shown in Figure~\ref{fig:useful-upvotes}
-- when the half-life is $0{.}5$ days ($90$ days), $42$\% ($24$\%) of upvotes lead to learning. 
\begin{figure}[t]
\centering
\includegraphics[width=0.35\textwidth]{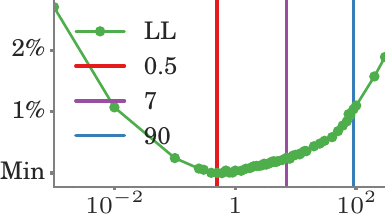}
\caption{Negative log-likelihood plotted for different values of kernel parameter (expressed as half-life in days). The $y$-axis shows the relative difference with respect to the
minimum value. The likelihood nearly plateaus ($\sim$$1\%$) for half-life between $0{.}5$ and $90$ days. The results we present are robust to parameter changing within the
range and we chose $7$ days as a representative value.}
\label{fig:ll-est}
\end{figure}

In the remaining sections, for ease of exposition, we set the kernel parameter such that the half-life of knowledge is 7 days (refer Figure~\ref{fig:ll-est}), however, the insights
obtained in the following sections are robust to changes in the kernel parameter.
\begin{figure*}[!!!t]
\centering
\subfloat[{\scriptsize Avg.\ learner (Avg.\ knowledge / contribution: 0.005)}]{\makebox[0.5\textwidth][c]{\includegraphics[width=0.5\textwidth]{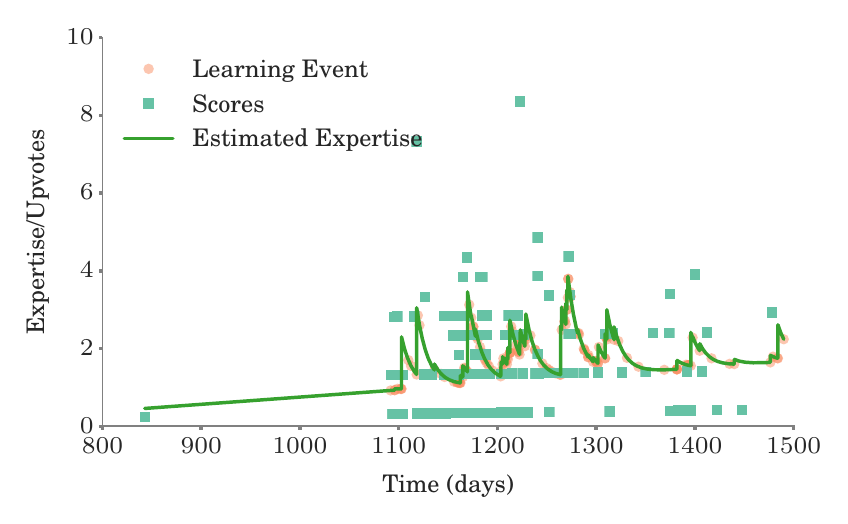} }\label{fig:learning-trajectory-real-2}}
\subfloat[{\scriptsize Expert: (Avg.\ knowledge / contribution: 0.034)}]{\makebox[0.5\textwidth][c]{\includegraphics[width=0.5\textwidth]{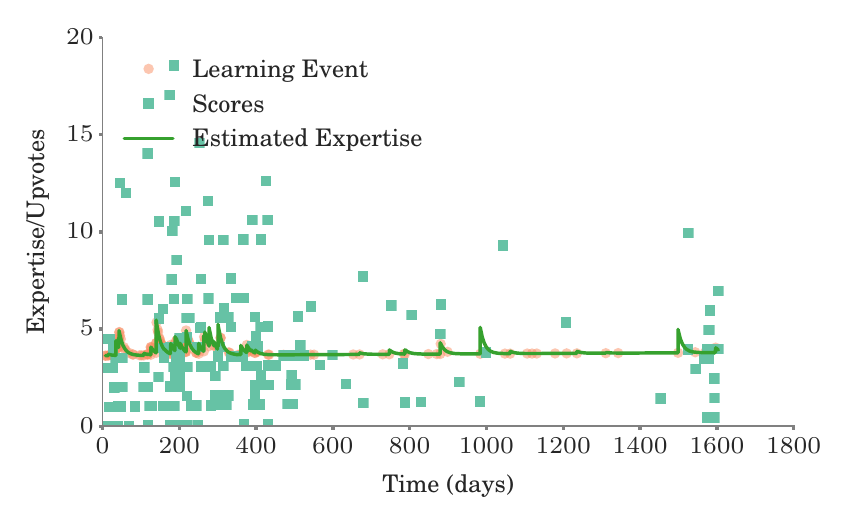}}\label{fig:learning-trajectory-real-3}}\\
\subfloat[{\scriptsize On-site learner (on-site learning: 55\%)}]{\makebox[0.5\textwidth][c]{\includegraphics[width=0.5\textwidth]{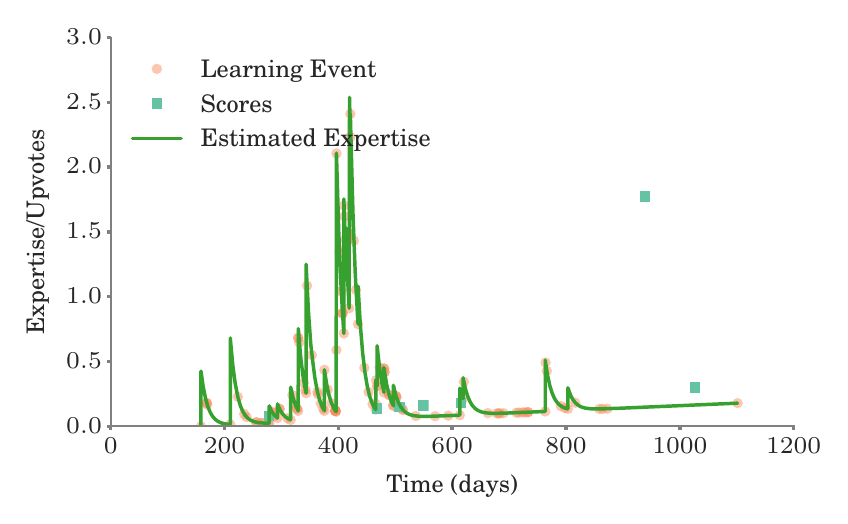}}\label{fig:learning-trajectory-real-1}}
\subfloat[{\scriptsize Off-site learner (on-site learning: 0.4\%)}]{\makebox[0.5\textwidth][c]{\includegraphics[width=0.5\textwidth]{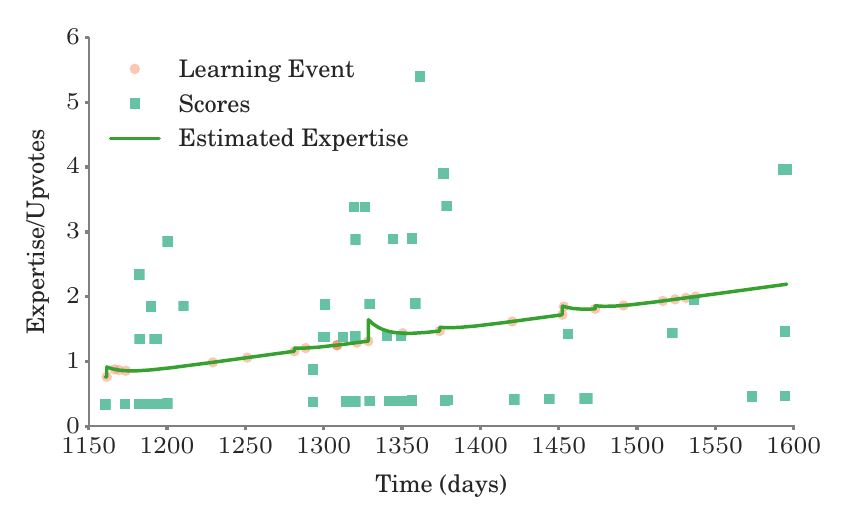}}\label{fig:learning-trajectory-real-4}}
\caption{Estimated learning trajectory for four characteristic Stack Overflow users.
The (average) learner (a) contributes answers with much less knowledge value than the
expert (b), \ie, $0.005$ vs $0.034$. The on-site learner (c) acquires $55$\% of her knowledge by learning from items in Stack Overflow
in contrast with the off-site learner (d), who only learns $0.4$\% of her knowledge by those means. Day $0$ in the plots is the date 2010-01-01.}\label{fig:learning-trajectory-real}
\end{figure*}

%
%
%
%

%
\xhdr{Types of learners}
Here, our goal is to better understand the type of learners that use crowd-learning sites as well as their characteristic properties.
To this aim, we start by visualizing the estimated learning trajectory for four different users --- an average learner, an on-site learner, an off-site learner and an expert ---
in Figure~\ref{fig:learning-trajectory-real}.
Each of the users exhibits different characteristic properties.
For example, the average learner contributes answers with much less knowledge value ($0.005$) than the expert ($0.034$), and the on-site learner acquires $55$\% of the knowledge by learning from items
in Stack Overflow in contrast with the off-site learner, who only learns $0.4$\% of the knowledge by those means.
%

\begin{figure}[t]
\centering
\subfloat[On-site and off-site learning]{\makebox[0.50\textwidth][c]{\includegraphics[width=0.40\textwidth]{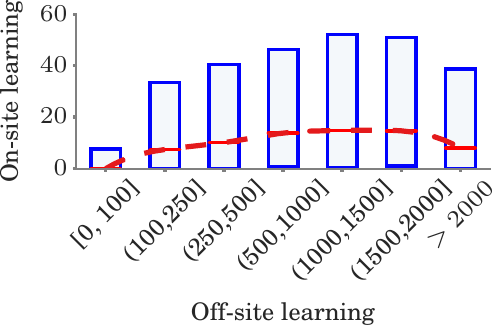}}\label{fig:in-site-off-site-learning}}
\subfloat[Starting and learned expertise]{\makebox[0.50\textwidth][c]{\includegraphics[width=0.40\textwidth]{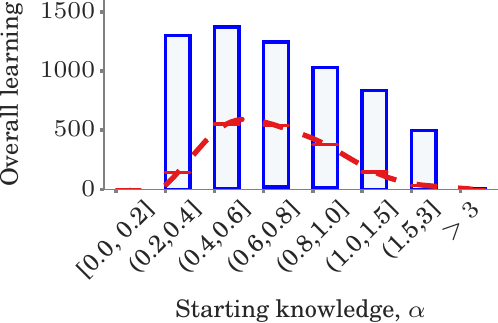}\label{fig:learning-baseline}}}
\caption{Behavior of learners in for tag \texttt{c\#}.
Panel (a) compares users'{} on-site and off-site learning in a box plot.
For $x\leq2000$, users achieving higher on-site learning also achieve higher off-site learning, however, over $x>2000$, users off-site learning
becomes more dominant.
Panel (b) shows users'{} overall learning against starting expertise in a box plot.
Users with very low or very high initial expertise, \ie, newbies and experts, tend to increase their knowledge the least, in contrast, users in the middle of the range tend to increase it the most.
In both panels, the limits of the boxes are the 25\%--75\% percentiles and the red dashed lines shows the median value.}\label{fig:properties-learners}
\end{figure}

Next, we investigate the interplay between on-site and off-site learning across all users. Here, given user $u$, we define on-site learning as the total expertise gathered by reading the knowledge items, {$\sum_{a \in \allTags} \sum_{q \in \Qut{u}{\Tall}}\int{\paramKwd{q}{a} \kappa_{\omega}(t)\, dt}$},
off-site learning as the expertise gathered outside Stack Overflow, {$\sum_{a \in \allTags} \int{\paramTrend{u}{a}}t\,dt$}, and overall learning as the sum of both.
{One can think of these quantities as the aggregate number of upvotes (\ie, score) users would have received on the site through either their on or off-site learning if they were posting answers at the same rate. Note that unlike the re\-pu\-ta\-tion on Stack Overflow, which is a measure of how much a user has effected others on the site, the on-site and off-site learning reflects how much a user has learned.}
Figure~\ref{fig:in-site-off-site-learning} compares users'{} on-site and off-site learning by means of a box plot.
For $x\leq2000$, users achieving higher on-site learning also achieve higher off-site learning, but over $x>2000$, off-site learning becomes more dominant.
Our results seem to indicate that quick learners rely less on  on-site learning, in relative terms.

Finally, we investigate the role that a user'{}s starting expertise plays on her overall learning over time by means of a box plot, shown in Figure~\ref{fig:learning-baseline}. Here, the $x$-axis corresponds to a user'{}s starting
expertise, $\alpha_u$, and the $y$-axis to her overall learning.
Interestingly, we find that users with very low or very high initial expertise, \ie, newbies and experts, tend to increase their knowledge the least, in contrast, users in the middle of the range tend to increase it the most. {This is in agreement with previous research, which indicated that
in presence of only positive reinforcement, the gain in expertise has a sigmoidal shape for learners, \ie, the newbies and experts increase their expertise at lower rates
than learners with medium levels of expertise~\cite{Leibowitz2010}.}

\begin{figure}[t]
\centering
\subfloat[Learned knowledge]{\makebox[0.30\textwidth][c]{\includegraphics[width=0.30\textwidth]{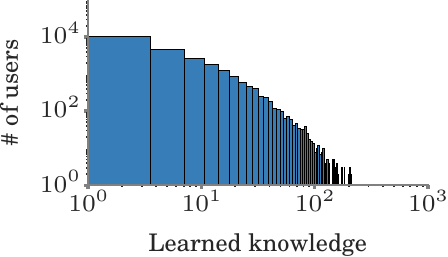}\label{fig:learners-distribution}}}
\subfloat[Contributed knowledge]{\makebox[0.30\textwidth][c]{\includegraphics[width=0.30\textwidth]{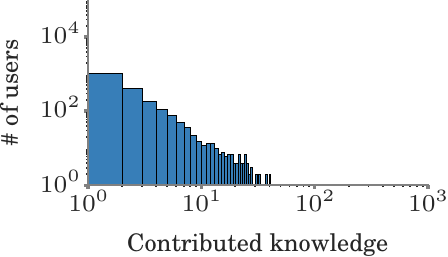}\label{fig:contributors-distribution}}}%
\subfloat[Avg.\ knowledge per contribution vs.\ learned knowledge]{\makebox[0.40\textwidth][c]{\includegraphics[width=0.35\textwidth]{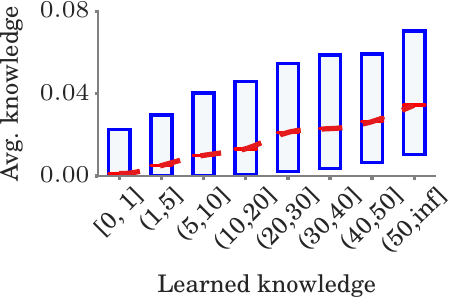}\label{fig:learners-contributors-scatter}}}
\caption{Learners vs contributors in Stack Overflow.
Panels (a) and (b) show the distribution of overall learned and contributed knowledge per user. The former follows a log-normal distribution {($\mu \approx 1.39, \sigma \approx 1.09$)}, while the latter follows a power-law {($\alpha \approx 2.26, x_{min} \approx 1.84$)}. {This shows that through the contributors are fewer in number than learners in absolute terms, they show much higher variability}.
Panel (c) shows a user'{}s average knowledge value per contribution against overall learned knowledge in a box plot. The red dotted line shows the median values and the box limits are the 25\%--75\% percentiles.
Interestingly, the users that learn more knowledge are also more proficient at producing high knowledge contributions.
}\label{fig:learners-contributors}
\end{figure}

\xhdr{Learners vs contributors}
A crowd-learning site is only useful if it has both learners and contributors. Here, we investigate two natural questions
that emerge in such context:
\begin{itemize}
\vspace{-1mm}
\item[I.] Are learners and contributors equally common?\vspace{-2mm}
\item[II.] Are more prolific learners better contributors?
\vspace{-1mm}
\end{itemize}
To answer the first question, we compute the distribution of learned and contributed knowledge per user.
Here, we estimate the knowledge value of each contribution (\ie, answer) in a knowledge item by dividing the total knowledge item value across contributions proportionally to their quality scores (upvotes).
Figures~\ref{fig:learners-distribution} and~\ref{fig:contributors-distribution} summarize the results. Although, in absolute numbers, there are more learners than contributors in our dataset, the amount of knowledge fed into the site by the contributors shows higher variability than the knowledge learned by users -- the distribution of contributed knowledge is fat tailed ($\alpha \approx 2.26$).

%
%

Next, we investigate the second question and assess whe\-ther more prolific learners are better contributors. To do so, we calculate the average knowledge value per contribution across users that have learned similar
amount of knowledge over time, \ie, sum of the knowledge value of all the knowledge items the user learned from, {$\sum_{a \in \allTags} \sum_{q \in \Qut{u}{\Tall}} \paramKwd{q}{a}$}. Fig\-ure~\ref{fig:learners-contributors-scatter}
shows that the users that learn more knowledge are also more proficient at producing high knowledge contributions.
In other words, our results suggest that ``\emph{by learning you will teach; by teaching you will learn}.''

%

\section{Discussion}\label{sec:discussion}

In this section, we take a step back and discuss the limitations of our model.
%
%
First, we remark that, due to the large number of parameters in the model, it is necessary to have access to {large amount of data} for our model estimation method to be accurate. However, this limitation can be overcome, to some extent, by linking expertise of a user across different platforms or sites (\eg, MOOCs), \ie, our model can easily assimilate traces available for the same user from those sites.

In our model, it is also crucial that the score reflects the true assessment of the knowledge content of the item and not of, say, the popularity of the contributor. In the case of Stack Overflow, which is a strict and self-regulated community, upvotes are
seldom granted to answers which do not address the question --- cases of serial upvoting are caught and remedied quickly which (mostly) prevents users from voting as a \emph{thank you} gesture. As a consequence, on Stack Overflow, upvotes
on answers are a good assessment of the quality of the posts. However, a sensible choice for scores in platforms or sites with milder self-regulation may be challenging.

Finally, the learning events also need to be chosen such that they are not conflated with other objectives the user may have on the website.
{On Stack Overflow, if a user only upvotes a question, it indicates that she relates with the problem but none of the answers (if any) provide a solution. However, upvoting an answer is evidence that the Q\&A pair \emph{taught} the user something.}

{These unique features and mechanisms afforded by Stack Overflow allow us to easily identify learning events and assessments. Finding similar features in a different social network would require careful reasoning and justification.}


\section{Conclusions}
In this paper, we proposed a probabilistic model of crowd\-lear\-ning, naturally designed to fit fine-grained learning and
contributing event data.
The key innovation of our model is modeling the evolution of users'{} expertise over time as a latent stochastic process,
driven by both off-site and on-site learning.
Then, we developed a scalable estimation method to fit the model parameters from millions of recorded learning events
and contributions.
Finally, we applied our model to a large set of learning and contributing events from Stack Overflow and found several
interesting insights.
For example, items with high knowledge value are rare.
%
Newbies and experts acquire less knowledge than users in the middle range.
Prolific learners tend to be also proficient contributors that share knowledge with high knowledge value.

%
%
%

Our work also opens many interesting venues for future work.
For example, natural follow-ups to potentially improve the expressiveness of our modeling framework include:

\vspace{1mm}
\begin{compactitem}
\item[1.] Consider more complex off-site learning trends, \eg, isotonic regression~\cite{kakade2011efficient} or exponential/power-law~\cite{heathcote2000power}.
\item[2.] Allow for a knowledge item to have different knowledge values per user by considering a knowledge distribution per item,
and use Bayesian inference~\cite{murphy2012machine} to learn the model parameters. 
\item[3.] Perform a non-parametric estimation of the kernels that model the users'{} forgetting process. We expect this to allow clustering of knowledge items which provide short-lived and long-lasting knowledge.
\item[4.] Incorporate incentives mechanisms such as badges, which are often used in crowdlearning sites~\cite{anderson2013steering} and MOOCs~\cite{anderson2014engaging}.
%
%
%
\end{compactitem}
\vspace{1mm}
One of the key modeling ideas behind our framework is realizing that users'{} contributions can be viewed as noisy discrete
samples of the users'{} expertise at points localized (non uniformly) in time.
We could generalize this idea to any type of event data and derive sampling theorems and conditions under
which an underlying general continuous signal of interest (be it user'{}s expertise, opinion, or wealth) can be recovered from
event data with provable guarantees.
Finally, we experimented with data gathered exclusively from Stack Overflow. It would be interesting to apply
our model to Stack Exchange at large, to other questions and answers websites (\eg, AskReddit),
microblogging platforms (\eg, Twitter), social networking sites (\eg, Pinterest), or even offline crowdlearning networks (\eg, citation network).

\xhdr{Acknowledgements} The authors would like to thank Sam Brand from Stack Overflow for providing data to make this work possible.


\vfill\eject
\small
\bibliographystyle{abbrv}
\bibliography{refs}  
\end{document}